\documentclass[12pt,a4paper]{article}
\usepackage{latexsym}
\usepackage{amsmath}
\usepackage{amsfonts}
\usepackage{amsbsy}
\usepackage{amssymb}
\usepackage{epsfig}
\usepackage{mathrsfs}
\usepackage{epsfig}
\usepackage{psfrag}
\usepackage{bbm}

\numberwithin{equation}{section}

\title{On knottings in the physical Hilbert space of LQG as given by the EPRL model}
\author{Benjamin Bahr$^{1,2}$\\[5pt]
\small $^1$ DAMTP, University of Cambridge,\\
\small  Wilberforce Road, Cambridge CB3 0WA, UK \\
\small   $^2$ MPI f. Gravitational Physics, Albert Einstein Institute,\\
 \small Am M\"uhlenberg 1, D-14476 Potsdam, Germany }

\begin{document}

\maketitle

\begin{abstract}
\noindent We consider the EPRL spin foam amplitude for arbitrary embedded two-complexes. Choosing a definition of the face- and edge amplitudes which lead to spin foam amplitudes invariant under trivial subdivisions, we investigate invariance properties of the amplitude under consistent deformations, which are deformations of the embedded two-complex where faces are allowed to pass through each other in a controlled way. Using this surprising invariance, we are able to show that in the physical Hilbert space as defined by the sum over all spin foams contains no knotting classes of graphs anymore.
\end{abstract}


\section{Introduction}

\begin{figure}[hbt!]
    \begin{center}
	\psfrag{g}{\small $\gamma$}	
	\psfrag{i1}{\small $\iota_1$}
	\psfrag{i2}{\small $\iota_2$}
	\psfrag{i3}{\small $\iota_3$}
	\psfrag{i4}{\small $\iota_4$}
	\psfrag{i5}{\small $\iota_5$}
	\psfrag{i6}{\small $\iota_6$}
	\psfrag{j1}{\small $k_1$}
	\psfrag{j2}{\small $k_2$}
	\psfrag{j3}{\small $k_3$}
	\psfrag{j4}{\small $k_4$}
	\psfrag{j5}{\small $k_5$}
	\psfrag{j6}{\small $k_6$}
	\psfrag{j7}{\small $k_7$}
	\psfrag{j8}{\small $k_8$}
	\psfrag{j9}{\small $k_9$}
	\psfrag{j10}{\small $k_{10}$}
\includegraphics[scale=0.3]{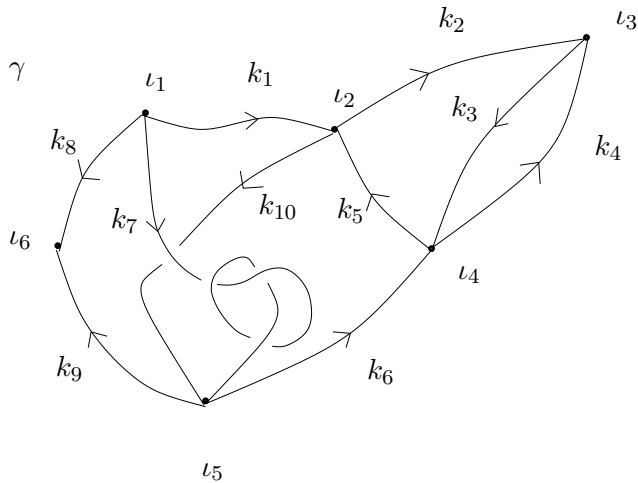}
    \caption{A spin network function consists of an embedded graph $\gamma\subset\Sigma$, spins $k_e$ along its edges and interwiners $\iota_v$ along its vertices.}\label{Fig:Graph}
    \end{center}
\end{figure}

Loop Quantum Gravity is an approach to canonical quantum gravity (see \cite{ALLMT, ROVELLI, INTRO} and references therein). The kinematical states $\psi$ of the theory are given by elements in an auxiliary Hilbert
space $\mathcal{H}_{\rm kin}$, and are considered to correspond to
quantized Cauchy data for the initial value problem of General
Relativity. An orthonormal basis for $\mathcal{H}_{\rm kin}$ is
given by the spin network functions, which carry a geometrical
interpretation given by the geometric operators associated to areas
and volumes. Quantum numbers associated to the states are graphs $\gamma$ embedded in the Cauchy surface $\Sigma$, spins $k$ (i.e. irreducible representations of $SU(2)$) along the edges of $\gamma$, together with $SU(2)$-intertwiners on its vertices (see figure \ref{Fig:Graph}).

Evolution of classical Cauchy data in General Relativity is subject
to constraints, which ensure the four-diffeomorphism invariance of
the theory. These constraints are divided into the (spatial)
diffeomorphism constraints $D_a$ and the Hamiltonian constraint $H$.
Correspondingly, the states $\psi$ in $\mathcal{H}_{\rm kin}$ are
not considered to be physical, but rather one is looking for states
$\psi_{\rm phys}$ which are solutions to the constraint equations
$\hat D_a\psi_{\rm phys}=\hat H\psi_{\rm phys}=0$.

In order to obtain the physical Hilbert space from the kinematical one, one usually employs a so-called "rigging map", which serves as a bona fide projector. Technically, this amounts to an anti-linear map
\begin{eqnarray}\label{Gl:RiggingMap}
\eta:\;\mathcal{D}_{\rm kin}\;\longrightarrow\;\mathcal{D}_{\rm kin}^*
\end{eqnarray}

\noindent which maps elements of a dense subspace $\mathcal{D}_{\rm kin}$ of $\mathcal{H}_{\rm kin}$ (which is usually taken to be the finite linear span of the spin network functions) into its algebraic dual. Hence states are mapped into distributions, and one equips the range of $\eta$ with an inner product via
\begin{eqnarray}\label{Gl:PhysicalInnerProduct}
\langle\,\eta[\phi]\,|\,\eta[\psi]\,\rangle_{\rm phys}\;:=\;\eta[\phi](\psi).
\end{eqnarray}

\noindent Generically the map $\eta$ has a non-trivial kernel, and hence one has to divide the zero space out of the range of  the inner product (\ref{Gl:PhysicalInnerProduct}) in order to make it positive definite. As a result, several different kinematical states $\psi$ will be mapped to the same physical state $\eta[\psi]$ by (\ref{Gl:RiggingMap}). This is definitely a desired feature, since, in the case of systems with constraints, states that are e.g. related by a change of gauge differ kinematically, but should be the same physically.\\[5pt]

In this article, we consider a specific proposal for the physical inner product (\ref{Gl:PhysicalInnerProduct}) in Loop Quantum Gravity, which mimics a path integral formulation for GR in the sense that it includes a sum over histories of spin networks. This idea goes back to Rovelli and Reisenberger (\cite{ROVREI, BAEZFOAM}, see \cite{PEREZ} for a review), and lies at the foundation of spin foam models as understood in the context of Loop Quantum Gravity.\footnote{The mathematical concept as abstract state sum model is much older (see e.g. literature in \cite{BC, PR}), but differs slightly from the way spin foams are understood in this article.}

Consider the $4d$ manifold $\mathcal{M}$ together with a foliation into $3d$ hypersurfaces $\mathcal{M}=\Sigma\times[0,1]$. Embed two states $\psi_i$ and $\psi_f$ into the initial and final hypersurface respectively. Now consider a two-complex $\kappa$ embedded in $\mathcal{M}$ that has $\psi_i$ and $\psi_f$ as boundary. This two-complex  (which is interpreted as history of the spin network state evolving from $\psi_i$ to $\psi_f$) is assigned the so-called "spin foam amplitude" $Z[\kappa]$, which is usually given in a local form
\begin{eqnarray}\label{Gl:SFAmplitude}
Z[\kappa]\;=\;\underbrace{\prod_{f}\mathcal{A}_f\prod_e\mathcal{A}_e\prod_v\mathcal{A}_v}_{\rm interior}\;\underbrace{\prod_{e}\mathcal{B}_e\prod_v\mathcal{B}_v}_{\rm boundary}
\end{eqnarray}

\noindent where a particular spin foam model is chosen by a specification of the amplitudes $\mathcal{A}_v$, $\mathcal{A}_e$ and $\mathcal{A}_f$ associated to the interior vertices, edges and faces of the two-complex, as well as the amplitudes $\mathcal{B}_v$ and $\mathcal{B}_e$ associated to the vertices and edges of the boundary spin networks $\psi_i$ and $\psi_f$.

The physical inner product, as given by the spin foam model, is then defined as the weighted sum over all two-complexes $\kappa$ that have $\psi_i$ and $\psi_f$ as boundary (omitting the $\eta$ from now on for reasons of brevity), i.e.
\begin{eqnarray}\label{Gl:PhysicalInnerProductWithFoams}
\langle\,\psi_f\,|\,\psi_i\,\rangle_{\rm phys}\;&=&\;\sum_{\kappa\,:\,\psi_i\;\stackrel{\kappa}{\longrightarrow}\;\psi_f}\;Z[\kappa]
\end{eqnarray}

\noindent This definition resembles a path-integral in the sense that it is a weighted sum over histories of spin networks, as proposed by Feynman. This particular form of the physical inner product as sum over embedded two-complexes has received increased interest recently (see in particular \cite{NEWLQG}).

Two of the spin foam models that are considered mostly today are the EPRL \cite{EPRL} and the FK model \cite{FK}. They have evolved out of a state-sum model constructed by Barrett and Crane \cite{BC}, and amount to specific choices of $\mathcal{A}_v$. It has been shown \cite{ASYMP1, ASYMP2} that, in the limit of large quantum numbers, the amplitudes $Z[\kappa]$, for $\kappa$ being the complex dual to a four-simplex, are asymptotically equal to the cosine of the action of Regge-discretised gravity for that simplex, i.e. one has
\begin{eqnarray}
Z[\kappa]\;\sim\,e^{i S_{\rm Regge}}\;+\; e^{-i S_{\rm Regge}},
\end{eqnarray}

\noindent which supports the belief that the sum (\ref{Gl:PhysicalInnerProductWithFoams}) is in some sense connected to a path integral for gravity.\footnote{It should furthermore be mentioned that there is a vast arsenal of technical tools for performing calculations in spin foam models built upon the definition of certain coherent states given in \cite{LS}, which also allows for an interpretation of the intertwiners in terms of four-dimensional geometry, see e.g. \cite{TWISTED, UN}.}

For all spin foam models that exist today however, the sum in (\ref{Gl:PhysicalInnerProductWithFoams}) diverges, since there are infinitely many two-complexes, even if one sums only over diffeomorphism-equivalence classes of foams.\footnote{Even for a fixed two-complex the sum over representation labels diverges. See e.g. \cite{FACE2}.} The task therefore still remains to make precise sense out of (\ref{Gl:PhysicalInnerProductWithFoams}).\\[5pt]

The scope of this article is to have a closer look at the symmetries of $Z[\kappa]$, i.e. to investigate on which properties of $\kappa$ the spin foam amplitude actually depends. The reason for this is two-fold. On the one hand, by finding a large symmetry group, one can restrict the sum over $\kappa$ to a sum over equivalence classes $[\kappa]$ under the symmetry, which makes it more likely to get hold of a mathematical meaningful concept of summation in (\ref{Gl:PhysicalInnerProductWithFoams}). On the other hand, we will be able to make certain statements about the actual size of the physical Hilbert space $\mathcal{H}_{\rm phys}$, i.e. to see which states in $\mathcal{H}_{\rm kin}$ are actually mapped to the same physical states. To be more specific, we will show that for the EPRL model, and any sensible definition of the sum (\ref{Gl:PhysicalInnerProductWithFoams}), states $\psi$ on graphs with the same combinatorics but different knotting classes will be mapped to the same physical state under (\ref{Gl:RiggingMap}). In other words, states in the physical Hilbert space will (at most) be labelled by combinatorial graphs, but will contain no knotting information.\\[5pt]

The plan for this article is as follows: We start by recalling the definition of the Euclidean EPRL spin foam model for arbitrary two-complexes in section \ref{Sec:Section2}. We moreover discuss a specific choice for the edge-, face- and boundary amplitudes. In section \ref{Sec:Deform} we define the consistent deformation of a spin foam $\kappa$, and show that the amplitude $Z[\kappa]$ is invariant under it. In section \ref{Sec:PhysHil} we describe a specific "unknotting" spin foam $\kappa_0$, compute its amplitude, and -- using the symmetry of the spin foam amplitudes demonstrated in the previous chapter -- show that the physical Hilbert space contains no knotting information of graphs anymore.

\section{The EPRL spin foam model a la KKL}\label{Sec:Section2}

\subsection{Labels on two-complexes}


In the following we recap in detail the definitions of the EPRL spin foam amplitudes, in order to make the reader familiar with our conventions, which keep close to those given in \cite{KKL}. The main difference to the more traditional ways of defining spin foams is that the two-complex in question is always regarded as embedded in a space-time manifold, and that at all times the orientation of the faces and edges of the two-complex are to be taken care of.\footnote{The orientation of the individual cells provides a natural pairing between the different tensors associated to the cells. In the numerous figures we often refrain from adding arrows indicating orientations whenever possible in order not to overburden the reader's eye.}

For the rest of the article, fix a $4d$ manifold $\mathcal{M}\sim\Sigma\times[0,1]$, where $\Sigma_i=\Sigma\times\{0\}$ and $\Sigma_f=\Sigma\times\{1\}$ are viewed as initial and final $3d$ hypersurface. By $\kappa$ we always denote a piecewise analytic\footnote{It is, up to this point, not clear which is the most general category of two-complexes one can work with. It is sure that piecewise-analytic works, just as piecewise analytic graphs in LQG are a convenient choice. Whereas LQG can also be defined with smooth graphs (which is much more involved, however, and spin network functions do not form a basis for 'smooth LQG'), it is not known if the EPRL model can be defined for, say, $CW$ complexes. } two-complex embedded in $\mathcal{M}$, such that the intersections of $\kappa$ with $\Sigma_{i,f}$ are piecewise analytic graphs embedded in either $\Sigma_{i,f}$. The two-complex $\kappa$ is endowed with an orientation, i.e. an orientation of all its $1$-cells ("edges") and $2$-cells ("faces")\footnote{The $0$-cells ("vertices") are equipped with the $+$-orientation by default.}. Furthermore, $\kappa$ is equipped with irreducible $Spin(4)$-representations $(j_f^+,j_f^-)$ (where the $j_f^\pm\in\frac{1}{2}\mathbb{N}$ are spins that arise from the equivalence $Spin(4)\simeq SU(2)\times SU(2)$) along its faces $f$, and intertwiners $\iota_e$ along its edges $e$. The convention is such that $\iota_e$ is an intertwiner between the tensor product of representations belonging to faces of which $e$ is a part of the boundary such that the orientation of $e$ and the induced orientation of $f$ coincide ("incoming faces"), and the tensor product of the faces that induce an orientation opposite to the one of $e$ ("outgoing faces").\\[5pt]

\begin{figure}[hbt!]
    \begin{center}
	\psfrag{e}{$e$}
    \psfrag{f1}{$f_1$}
    \psfrag{f2}{$f_2$}
    \psfrag{f3}{$f_3$}
\includegraphics[scale=0.85]{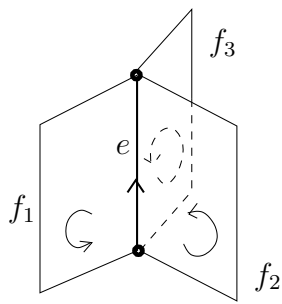}
    \caption{An edge $e$ with bordered by three faces. The intertwiner $\iota_e$ associated to it maps $\iota_e\;:\;V_{j_{f_1}^+,j_{f_1}^-}\;\longrightarrow\;V_{j_{f_2}^+,j_{f_2}^-}\otimes V_{j_{f_3}^+,j_{f_3}^-}$ and intertwines the action of $Spin(4)\simeq SU(2)\times SU(2)$ on both sides.}
    \end{center}
\end{figure}

\noindent Each intertwiner map

\begin{eqnarray}\label{Gl:Intertwiner}
\iota_e\;:\;\bigotimes_{f\text{ incoming}}V_{j_{f}^+,j_{f}^-}\;\longrightarrow\;\bigotimes_{f\text{ outgoing}}V_{j_{f}^+,j_{f}^-}
\end{eqnarray}

\noindent is to commute with the group action on either side of (\ref{Gl:Intertwiner}), i.e.
\begin{eqnarray}
\iota_e\circ\left(\bigotimes_{f\text{ incoming}}\rho_{j_{f}^+,j_{f}^-}(g)\right)\;=\;\left(\bigotimes_{f\text{ outgoing}}\rho_{j_{f}^+,j_{f}^-}(g)\right)\circ\iota_e
\end{eqnarray}

\noindent for all $g\in Spin(4)$, when $\rho_{j_f^+,j_f^-}$ is the irreducible representation on $V_{j_f^+,j_f^-}$ belonging to the face $f$.

\subsection{EPRL intertwiner}

\begin{figure}[hbt!]
\begin{center}
    \psfrag{d}{$\cdots$}
    \psfrag{i}{$\!\!\iota=\phi(\hat\iota)$}
    \psfrag{ii}{$\hat\iota$}
    \psfrag{k1}{$k_1$}
    \psfrag{k2}{$k_2$}
    \psfrag{kn}{$k_n$}
    \psfrag{j11}{$j_1^+$}
    \psfrag{j12}{$j_1^-$}
    \psfrag{j21}{$j_2^+$}
    \psfrag{j22}{$j_2^-$}
    \psfrag{jn1}{$j_n^+$}
    \psfrag{jn2}{$j_n^-$}
    \includegraphics[scale=0.9]{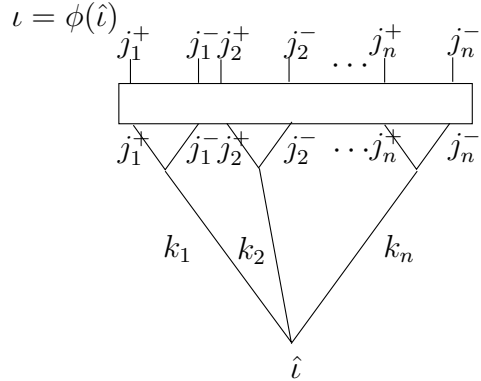}
    \caption{Diagrammatic representation of the map $\phi$. The lines carry $SU(2)$ representations, the box denotes integration over $SU(2)\times SU(2)$, and each vertex carries an $SU(2)$ intertwiner. All face are outgoing, so there is no additional index $out,\,in$ to distinguish them.}\label{Fig:Diagram}
\end{center}
\end{figure}

The EPRL model consists of a special choice of intertwiner (\ref{Gl:Intertwiner}). Equivalently, the model declares all amplitudes $\mathcal{A}_e$, $\mathcal{A}_f$, $\mathcal{A}_v$ to be equal to zero whenever the corresponding representations $j_f^{\pm}$ and intertwiners $\iota_e$ do not satisfy the following conditions for a fixed real $\gamma\neq\pm 1,0$:
\begin{itemize}
\item For each face $f$ there is a half-integer $k_f\in\frac{1}{2}\mathbb{N}$ such that
\begin{eqnarray}\label{Gl:EPRLrepresentations}
j_f^\pm\;=\;\frac{|1\pm \gamma|}{2}k_f
\end{eqnarray}
\item For each edge $e$ the associated intertwiner is the image of the following map $\phi$, which maps $SU(2)$ intertwiners to $SU(2)\times SU(2)$-intertwiners:
    \begin{eqnarray}\nonumber
    \phi\;:\;\bigotimes_{f\text{ incoming}}V_{k_f}&\otimes&\bigotimes_{f\text{ outgoing}}V_{k_f}^\dag\;\longrightarrow\;\bigotimes_{f\text{ incoming}}V_{j_{f}^+,j_{f}^-}\otimes\bigotimes_{f\text{ outgoing}}V_{j_{f}^+,j_{f}^-}^\dag\\[5pt]\nonumber
    \phi(\hat\iota)_{m_1^+m_1^-,\ldots m_k^+m_k^-}{}^{n_1^+n_1^-,\ldots n_l^+n_l^-}&\;=\;&\int dh^+dh^-\,\pi_{j_{in,1}^+}(h^+)_{m_1^+\tilde{m}_1^+}\ldots \pi_{j_{in, k}^+}(h^+)_{m_k^+\tilde{m}_k^+}\\[5pt]\nonumber
    &&\;\times\;\pi_{j_{in, 1}^-}(h^-)_{m_1^-\tilde{m}_1^-}\ldots \pi_{j_{in, k}^-}(h^-)_{m_k^-\tilde{m}_k^-}\\[5pt]\label{Gl:Boost}
    &&\;\times\;\pi_{j_{out,1}^+}((h^+)^{-1})_{\tilde{n}_1^+n_1^+}\ldots \pi_{j_{out,k}^+}((h^+)^{-1})_{\tilde{n}_k^+n_k^+}\\[5pt]\nonumber
    &&\;\times\;\pi_{j_{out,l}^-}((h^-)^{-1})_{\tilde{n}_l^-n_l^-}\ldots \pi_{j_{out, k}^-}((h^-)^{-1})_{\tilde{n}_k^-n_k^-}\\[5pt]\nonumber
    &&\;\times\;C_{j_{in,1}^+\tilde{m}_1^+j_{in,1}^-\tilde{m}_1^-}^{k_{in,1}p_1}\cdots C_{j_{in,k}^+\tilde{m}_k^+j_{in,k}^-\tilde{m}_k^-}^{k_{in,k}p_k}\\[5pt]\nonumber
    &&\;\times\; C_{j_{out,1}^+\tilde{n}_1^+j_{out,1}^-\tilde{n}_1^-}^{k_{out,1}q_1}\cdots C_{j_{out,l}^+\tilde{n}_l^+j_{out,l}^-\tilde{n}_l^-}^{k_{out,l}q_l}\\[5pt]\nonumber
    &&\;\times\;\hat{\iota}_{p_1\cdots p_k}{}^{q_1\cdots q_l}
    \end{eqnarray}
\noindent I.e.  $\iota_e=\phi(\hat\iota_e)$ for some $SU(2)$-intertwiner $\hat\iota_e$. Here $2j_{x,i}^\pm=|1\pm\gamma|k_{x,i}$ for $x=\{in,out\}$ are the $SU(2)\times SU(2)$ representations to the ingoing and outgoing faces. The $C_{j_1m_1j_2m_2}^{JM}$ are the usual Clebsch-Gordon coefficients  \cite{AMQG}. The complicated-looking formula (\ref{Gl:Boost}) can be conveniently summarised with the help of the diagrammatic calculus given in figure \ref{Fig:Diagram}\footnote{Which is, strictly speaking, only true if all faces are outgoing. See e.g. \cite{ASYMP1} for an introduction to the diagrammatic calculus, which allows to avoid writing down expressions such as (\ref{Gl:Boost}).}
\end{itemize}

\noindent It should be noted that the map (\ref{Gl:Boost}) is often referred to as "boost" in the literature. Since it is an isomorphism between the space of $SU(2)$-intertwiners and EPRL-intertwiners (i.e. those $SU(2)\times SU(2)$-intertwiners which satisfy the simplicity constraints \cite{EPRL}), it provides an explicit one-to-one correspondence between data on the two-complex, and data associated to boundary states known from LQG.\footnote{ It has been shown \cite{NOISO} that $\phi$ is no isometry, which has lead to discussion about the correct way in which to sum over the intertwiner data in (\ref{Gl:PhysicalInnerProductWithFoams}). Since we perform calculations with  single amplitudes only and leave the explicit form of the sum over spin foams in (\ref{Gl:PhysicalInnerProductWithFoams}) open, the results of this article are independent of this question.}\\[5pt]

As indicated, the spin foam labels $j_f^\pm$, $\iota_e$ induce - via the above maps - spin network labels $k_e$, $\hat\iota_v$ on the boundary graphs $\psi_i$, $\psi_f$, which can therefore be seen as elements in the kinematical Hilbert space $\mathcal{H}_{\rm kin}=L^2(\overline{\mathcal{A}},d\mu_{AL})$ of Loop Quantum Gravity. A spin foam $\kappa$ with these labels is denoted as
\begin{eqnarray}
\psi_i\;\stackrel{\kappa}{\longrightarrow}\;\psi_f
\end{eqnarray}

\noindent and is interpreted as the evolution of spin network functions starting as $\psi_i$ and ending as $\psi_f$.

\subsection{The vertex amplitude}


In the following we present the definition of the vertex amplitude $\mathcal{A}_v$ for a vertex $v$ in the two-complex $\kappa$, which is a function of the representations $j^\pm_f$ and intertwiners $\iota_e$ associated to faces and edges of $\kappa$. The data is supposed to satisfy the conditions (\ref{Gl:EPRLrepresentations}) and (\ref{Gl:Boost}).

Embed in $\mathcal{M}$ a small three-sphere, centered around the vertex $v$. The sphere $S^3$ should be placed such that it intersects each edge $e$ which meets $v$ at exactly one vertex $v_e$, and each face $f$ (which spans between two edges $e_1$, $e_2$) meeting $v$ at exactly one edge $e_f$ between the vertices $v_{e_1}$, $v_{e_2}$. Hence every vertex provides a dimensional reduction of the neighbourhood of $v$.

\begin{figure}[hbt!]
\begin{center}
    \psfrag{r1}{$j_1^\pm$}
    \psfrag{r2}{$j_2^\pm$}
    \psfrag{r3}{$j_3^\pm$}
    \psfrag{i1}{${\iota_e}$}
    \psfrag{S}{$S^3$}
    \psfrag{g}{$\gamma_v$}
    \psfrag{v}{$v$}
    \psfrag{k}{$\kappa$}
    \includegraphics[scale=0.9]{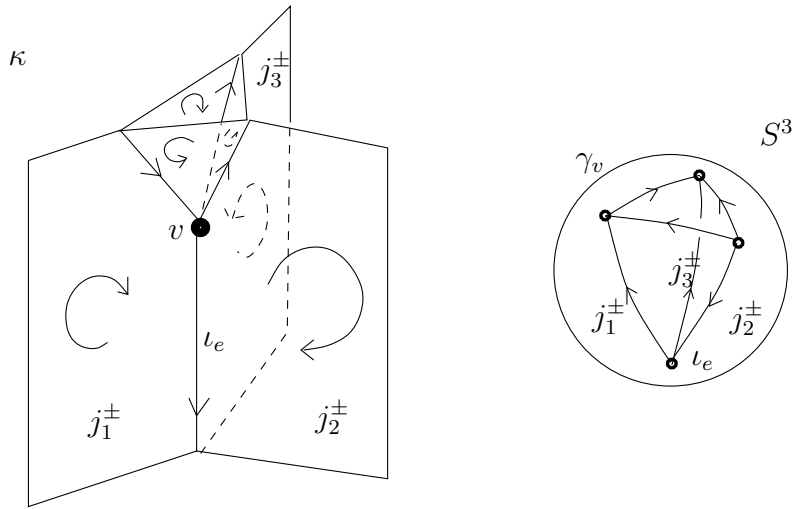}
    \caption{A vertex $v$ in a spin foam $\kappa$, and the associated vertex function $T_{\gamma_v,j^\pm_f,\iota_e}$, which is a $SU(2)\times SU(2)$-network function on the vertex graph $\gamma_v$ on a three sphere $S^3$, which results as a dimensional reduction of the neighbourhood of $v$. The edges and vertices of $\gamma_v$ correspond to faces and edges in $\kappa$ touching $v$.}\label{Fig:Vertex}
    \end{center}
\end{figure}

As a result, the intersection of the two-complex $\kappa$ with the sphere results in a graph $\gamma_v$, which consists of the edges and vertices $e_f$, $v_e$. Furthermore, the data $j_f^\pm$, $\iota_e$ determine data on the graph $\gamma_v$ in the following way: To every edge $e_f$ assign the orientation coinciding with the one of $f$, as well as the $Spin(4)$-representation $j_f^\pm$. To the vertices $v_e$ assign the intertwiner $\iota_e$, when $e$ is starting at the vertex $v$, and $\iota^{\dag}_e$ when $e$ is ending at $v$. It is not hard to see that this results in a $Spin(4)$-network function $T_{\gamma_v,j_f^\pm,\iota_e}$ based on $\gamma_v$. It is termed the \emph{vertex function}, and the vertex amplitude is defined as the evaluation on the flat connection
\begin{eqnarray}\label{Gl:VertexAmplitude}
\mathcal{A}_v\;&:=&\;T_{\gamma_v,j_f^\pm,\iota_e}(\mathbbm{1})\\[5pt]\nonumber
&=&\;\left(\prod_f\delta_{b_f}^{a_f}\right)\left(\prod_{e}(\iota_e)_{a_1a_2\ldots}{}^{b_1b_2\ldots}\right)
\end{eqnarray}

\noindent Effectively the amplitude is computed by contracting the indices of the EPRL intertwiners $\iota_e$ according to the combinatorics of the graph $\gamma_v$.

\subsection{Face-, edge- and boundary amplitudes}

In this section we consider a specific choice for the amplitudes that are assigned to faces $\mathcal{A}_f$ and edges $\mathcal{A}_e$ within (\ref{Gl:SFAmplitude}). We show how these amplitudes are derived by requiring the spin foam amplitude to be invariant under trivial subdivisions of the two-complex, which do not change its topology, as well as by requiring a functorial property of the spin foam amplitude. This particular choice of amplitudes has been considered before and is the most usual one (see e.g. \cite{FACE2}). It is however not the only one, see e.g. \cite{FACE} for arguments for different amplitudes.

\begin{center}
\begin{figure}[hbt!]
    \psfrag{j}{$j_f$}
    \psfrag{e}{${\iota_e}$}
    \psfrag{=}{$=$}
    \psfrag{E1}{$\iota_e={\rm id},\;\mathcal{A}_{f}\stackrel{!}{=}\mathcal{A}_f^2\mathcal{A}_e$}
    \psfrag{E2}{$\Rightarrow\,\mathcal{A}_f=\frac{1}{\mathcal{A}_e}=(2j_f^++1)(2j_f^-+1)$}
    \includegraphics[scale=0.6]{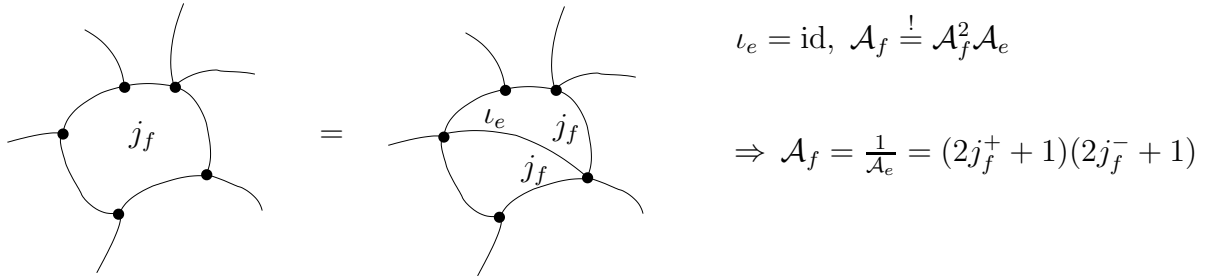}
    \caption{Trivial subdivision of a face with an edge}\label{Fig:SubdivisionFace}
\end{figure}
\end{center}

The spin foams described in this article are thought of as histories of spin networks of Loop Quantum Gravity. These spin networks themselves are invariant under trivial subdivisions, i.e. by dissecting an edge $e$ of a graph $\gamma$ with spin $k$ into two edges $e=e_1\circ e_2$ with spin $k$, and placing at the newly appeared vertex $v$ the identity intertwiner $\hat\iota_v=\mathbbm{1}_{V_{k}}$ (See figure \ref{Fig:SubdivisionFace}). This invariance suggests that one might desire a similar set of invariances to also hold for the spin foam amplitudes.

\begin{center}
\begin{figure}[hbt!]
    \psfrag{i}{$\iota_e$}
    \psfrag{v}{$v$}
    \psfrag{=}{$=$}
    \psfrag{E1}{$\mathcal{A}_e\stackrel{!}{=}\mathcal{A}_e^2\mathcal{A}_v$}
    \psfrag{E2}{$\Rightarrow\,\mathcal{A}_e=\frac{1}{\mathcal{A}_v}=\frac{1}{{\rm tr} (\iota_e\iota_e^{\dag})}$}
    \includegraphics[scale=0.6]{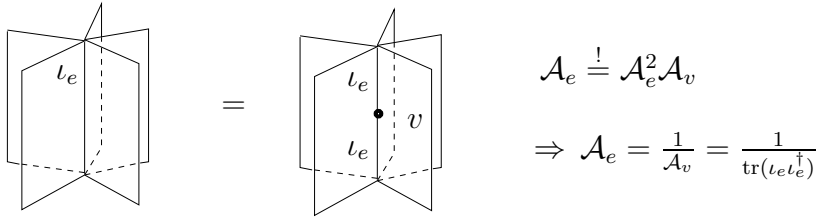}
    \caption{Trivial subdivision of an edge with a vertex.}\label{Fig:SubdivisionEdge}
\end{figure}
\end{center}

One way of subdividing a two-complex without changing its topology is by subdividing an edge $e=e_1\circ e_2$ by adding another vertex $v=e_1\cap e_2$.\footnote{Note that one doesn't have to subdivide the faces touching $e$ as well.} See figure \ref{Fig:SubdivisionEdge}.

The two new edges obtain the same orientation and intertwiner, i.e. $\iota_{e_1}=\iota_{e_2}:=\iota_e$. Invariance of $Z[\kappa]$ under this subdivision leads to the vertex- and edge amplitudes having to satisfy $\mathcal{A}_e=\frac{1}{\mathcal{A}_v}$. The vertex amplitude $\mathcal{A}_v$ can be evaluated using (\ref{Gl:VertexAmplitude}), realising that $\gamma_v$ is an $n$-bridge graph, where $n$ is the number of faces meeting $e$. It is given by $\mathcal{A}_v={\rm tr}(\iota_e\iota_e^\dag)$ (assuming that all faces are oriented along with $e$), leading to the edge amplitude
\begin{eqnarray}\label{Gl:EdgeAndVertexIdentity}
\mathcal{A}_e\;=\;\frac{1}{{\rm tr}(\iota_e\iota_e^\dag)}.
\end{eqnarray}

\noindent Another way of trivially subdividing the two-complex without changing its topology is by dissecting a face $f$ with representation $(j_f^+,j_f^-)$ into two faces $f_1$, $f_2$, each with the same orientation and representations $j_{f_1}^\pm=j_{f_2}^\pm=j_f^\pm$.\footnote{See also \cite{FACE2} for a more refined argument for this kind of invariance, considering the anomalies in the sum (\ref{Gl:PhysicalInnerProductWithFoams}).} The intertwiner $\iota_e$ on the newly appeared edge $e=f_1\cap f_2$ is chosen to be the identity intertwiner $\iota_e=\mathbbm{1}_{V_{j_f^+,j_f^-}}$ (see figure \ref{Fig:SubdivisionFace}).

If one demands invariance of the spin foam amplitude $Z[\kappa]$ under a trivial subdivision of a face, the edge- and face amplitudes obviously have, by (\ref{Gl:SFAmplitude}), to satisfy

\begin{eqnarray}\label{Gl:FaceAndEdgeIdentity}
\mathcal{A}_f=\frac{1}{\mathcal{A}_e},
\end{eqnarray}

\noindent where here $e$ is supplied with the identity intertwiner. With (\ref{Gl:EdgeAndVertexIdentity}) and (\ref{Gl:FaceAndEdgeIdentity}) one arrives at an expression for the face amplitude

\begin{eqnarray}
\mathcal{A}_f\;=\;(2j_f^++1)(2j_f^-+1).
\end{eqnarray}

\noindent It should be noted that, unlike the invariance under trivial subdivision of faces, the invariance of $Z[\kappa]$ under subdivision of an edge does not follow from the invariance of the spin networks under subdivision of an edge. We feel, however, that it is a natural invariance to demand, not only because this way one is able to fix both face- and edge-amplitudes in terms of the vertex amplitude, but also because the history of the spin network is not changed by this move, because the topology of the two-complex remains the same. It therefore seems natural that the spin foam amplitude, which contains information about the dynamics, remains the same as well.

It should also be noted that the two moves described above generate all subdivisions of the two-complex which do not change the history of the spin network, i.e. which are trivial subdivisions from the point of view of the dynamics.

Note finally that the invariance of the Spin foam amplitude under subdivision of an edge results in a formula for the amplitude which is homogenous in the scaling of the intertwiner of degree zero. The choice (\ref{Gl:EdgeAndVertexIdentity}) is equivalent to a normalisation of the vertex amplitudes, and can therefore be absorbed into a redefinition of $\mathcal{A}_v$, see e.g. \cite{ASYMP1} (which we, however, refrain from doing here). The choice of the face amplitude also seems natural from the point of view of BF theory. See however \cite{FACE} for a different amplitude, which in particular has the advantage of reducing the degree of divergence in the sum over spin foams (\ref{Gl:PhysicalInnerProductWithFoams}).


\subsection{Boundary amplitudes}

In the following, we describe the boundary amplitudes $\mathcal{B}_e$, $\mathcal{B}_v$, which appear in the spin foam amplitude (\ref{Gl:SFAmplitude}). The two-complex $\kappa$, which is embedded in $\mathcal{M}=\Sigma\times[0,1]$, induces boundary graphs $\gamma_i=\kappa\cap\Sigma\times\{0\}$ and $\gamma_f=\kappa\cap\Sigma\times\{1\}$. The spin foams $\kappa$ we are considering are all such that the initial and final dynamics of the spin network, which is described by the two-complex, is trivial. In more technical terms this means that there is an $\epsilon>0$ such that
\begin{eqnarray}\label{Gl:BoundaryCondition}
\begin{array}{rcl}
\kappa\cap\big(\Sigma\times[0,\epsilon]\big)\;&=&\;\gamma_i\times[0,\epsilon]\\[5pt]
\kappa\cap\big(\Sigma\times[1-\epsilon,1]\big)\;&=&\;\gamma_f\times[1-\epsilon,1]
\end{array}
\end{eqnarray}

\noindent The equality signs in (\ref{Gl:BoundaryCondition}) are meant to denote diffeomorphic equivalence. This condition on the topology of the two-complex is not as restrictive as one might initially think. In fact it is generic and amounts to the condition that cutting of a two-complex is only allowed when not cutting directly through an inner vertex.

The condition (\ref{Gl:BoundaryCondition}) also ensures that there is no nontrivial dynamics which is happening exactly at the initial or final boundary. In particular, let $\kappa_1$ and $\kappa_2$ be two two-complexes with $\gamma_{1,f}=\gamma_{2,i}=\gamma$, then there is an obvious way to concatenate the two two-complexes by glueing them together at $\gamma$. Then (\ref{Gl:BoundaryCondition}) ensures that in this process no nontrivial vertices appear. The result of this glueing process is denoted by $\kappa:=\kappa_1 \kappa_2$.

Since the amplitude $Z[\kappa]$ can be thought of $\exp[i S_{\rm EH}[g_{\mu\nu}]]$, where $\kappa,j_f^\pm,\iota_e$ is interpreted as an appropriate discretisation of $g_{\mu\nu}$ and the (Euclidean) Einstein-Hilbert is additive if the appropriate boundary conditions are included, it appears natural that one should demand
\begin{eqnarray}\label{Gl:MultiplicativeAmplitude}
Z[\kappa_1\kappa_2]\;=\;Z[\kappa_1]\,Z[\kappa_2]
\end{eqnarray}

\noindent and this can be achieved, as we will show now, by an according choice for the boundary amplitudes
$\mathcal{B}_e$, $\mathcal{B}_v$ in (\ref{Gl:SFAmplitude}).

There are several possibilities to choose the boundary amplitudes to ensure (\ref{Gl:MultiplicativeAmplitude}). One immediate choice is the following: For any vertex $v$ in the boundary, denote the edge that ends at $v$ by $e_v$. Similarly, denote for every boundary edge $e$ the face in $\kappa$ that ends in it by $f_e$.\footnote{Note that this is only well-defined if the topology of the two-complex satisfies (\ref{Gl:BoundaryCondition}).} Then, if one defines
\begin{eqnarray}\label{Gl:BoundaryAmplitudes}
\mathcal{B}_v\;:=\;\big(\mathcal{A}_{e_v}\big)^{-\alpha}\,\qquad\,\mathcal{B}_e\;:=\;\big(\mathcal{A}_{f_e}\big)^{\alpha-1}
\end{eqnarray}

\noindent for an $\alpha\in[0,1]$, it is straightforward to verify (\ref{Gl:MultiplicativeAmplitude}). Note that for nonzero intertwiner $\iota_e$, the amplitudes $\mathcal{A}_e$, $\mathcal{A}_f$ are always positive, hence (\ref{Gl:BoundaryAmplitudes}) is well-defined.

For the rest of the article, we adopt the symmetric choice $\alpha=\frac{1}{2}$, i.e.
\begin{eqnarray}\label{Gl:BoundaryAmplitudeSymmetric}
\mathcal{B}_v\;:=\;\frac{1}{\sqrt{\mathcal{A}_{e_v}}}\,\qquad\,\mathcal{B}_e\;:=\;\frac{1}{\sqrt{\mathcal{A}_{f_e}}}
\end{eqnarray}

\noindent which in particular leads to a more symmetric behaviour of the amplitudes under time-reversal.

\section{Consistent deformations}\label{Sec:Deform}

In this section we will demonstrate that the spin foam amplitudes $Z[\kappa]$ defined by the EPRL model possess a large symmetry, if one adapts the choices for the amplitudes from the last chapter. This symmetry will help us in proving the core statements about the properties of the physical inner product (\ref{Gl:PhysicalInnerProductWithFoams}) using the EPRL amplitude.

The first, trivially notable symmetry is that of ${\rm Diff}(\mathcal{M})^+$, i.e. the orientation-preserving diffeomorphisms of $\mathcal{M}$, since the actual value of the amplitude is only constructed using the combinatorial information of the embedding $\kappa\subset\mathcal{M}$, e.g. which faces are connected to which edges. The change of the amplitude under an orientation-changing diffeomorphism is discussed in \cite{KKL}.

In this chapter we will, however, turn to a larger invariance of $Z[\kappa]$ than just diffeomorphisms. Provided we choose the amplitudes in (\ref{Gl:SFAmplitude}) according to the last chapter, we will show that the spin foam amplitude is invariant under something which we term a \emph{consistent deformation}, which does not only involve deformations of two-complexes which are induced by diffeomorphisms of the embedding manifold $\mathcal{M}$, but in which e.g. some faces of $\kappa$ are allowed to pass through each other in a controlled way. Since by letting the surfaces of $\kappa$ intersect each other, one is, of course, creating new vertices, edges and faces. One therefore has to be careful what happens to e.g. the intertwiners in this case.\\[5pt]

We will have a look at two different "moves" that incorporate non-diffeomorphic deformations of the spin foam, and which leave the spin foam amplitude invariant. We will subsequently discuss the kind of deformations that are generated by these moves and trivial subdivisions of the spin foam.

\subsection{Pulling vertices apart}

\begin{figure}[hbt!]
\begin{center}
    \psfrag{v}{$v$}
    \psfrag{v1}{$v_1$}
    \psfrag{v2}{$v_2$}
    \psfrag{g}{$\gamma_v$}
    \psfrag{g1}{$\gamma_{v_1}$}
    \psfrag{g2}{$\gamma_{v_2}$}
    \psfrag{R}{$\Rightarrow$}
    \includegraphics[scale=0.7]{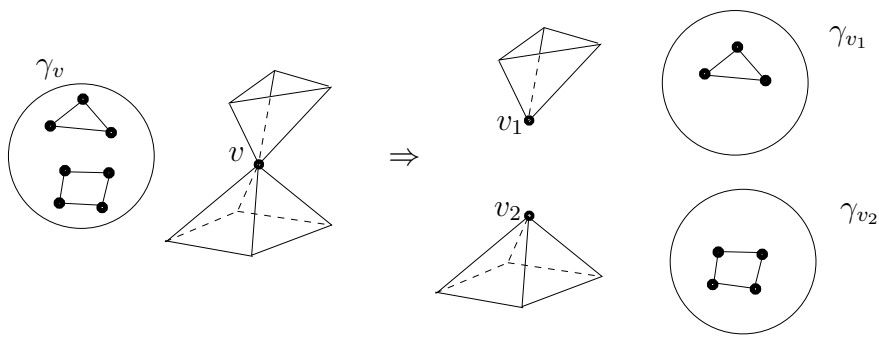}
    \caption{A vertex $v$ in a spin foam which has a disconnected vertex graph $\gamma_v$ can be pulled apart into two vertices $v_1$ and $v_2$}\label{Fig:PullVertices}
\end{center}
\end{figure}

The first move we are going to consider will be able to subdivide an inner vertex into two vertices. Consider an inner vertex $v$ of a spin foam that has the property that the corresponding vertex spin network $\gamma_v$ is disconnected in $S^3$, i.e. $\gamma_v=\gamma_{1}\cup\gamma_2$. Let furthermore $\gamma_1$ and $\gamma_2$ not be linked, i.e. one can, by a diffeomorphism of $S^3$, move e.g. $\gamma_1$ to the upper hemisphere and $\gamma_2$ to the lower hemisphere of $S^3$. If this can be achieved, then one can place a three-dimensional hypersurface $H$ in a neighbourhood of $v\in\mathcal{M}$ which intersects $\kappa$ exactly at $v$ (and the embedded sphere $S^3$ at the equator, see figure \ref{Fig:PullVertices}).

\noindent It is clear that $H$ divides a neighbourhood of $v$ of $\kappa$ into two parts which only intersect at $v$. The deformation we are defining consists of separating the two parts.\footnote{This separation can be achieved by e.g. defining a homotopy of the two-complex in the piecewise analytic category.} Of course, as soon as $\kappa$ is separated that way, the combinatorics of the resulting $\kappa'$ is different: $\kappa'$ has two vertices $v_1$, $v_2$ where $\kappa$ had only one, namely $v$. The number of edges and faces under this deformation stay the same, as well as the labels on them. The amplitude $Z[\kappa']$ can hence be computed, and it is not hard to see that
\begin{eqnarray}
\mathcal{A}_{v_1}\mathcal{A}_{v_2}\;=\;\mathcal{A}_v
\end{eqnarray}

\noindent since the evaluation of a disconnected spin network is just the product of the evaluation of each of its components. It follows immediately that
\begin{eqnarray}
Z[\kappa]\;=\;Z[\kappa'].
\end{eqnarray}

\noindent This "pulling apart" of the vertex $v$ was only possible since $\gamma_v$ was disconnected, so that its vertex amplitude $\mathcal{A}_v$ factorises. Of course, the move can be reversed by choosing two vertices $v_1$, $v_2$ in $\kappa$\footnote{Which can, in $\mathcal{M}$, be moved close to each other by a diffeomorphism of $\mathcal{M}$.} and moving them close to each other, so that they eventually overlap. This deformations, of course, needs to be done such that no other part of $\kappa$ intersects with another part. The resulting vertex $v$ has a vertex graph $\gamma_v$ which is of obviously disconnected so that its vertex amplitude is the product of the amplitudes of $v_1$ and $v_2$.

Note that this move is possible independently of the labels on $\kappa$. Whether this move can be performed on a vertex or not depends only on the embedding and topology of the two-complex $\kappa$. Furthermore, it is noteworthy that the move can be generalised to vertices $v$ whose vertex graph $\gamma_v$ have more than two connected components. Then $v$ can, by successive application of the above move, be separated into as many vertices as there are connected components in $\gamma_v$. The resulting spin foam $\kappa'$ has unchanged amplitude, and $\kappa'$ does not (modulo diffeomorphism) depend on the order in which the move above is applied.

\subsection{Pulling edges apart}

The second move (together with its inverse) which we are going to present is the higher-dimensional equivalent of pulling the vertices $v$ apart which have a factorising vertex amplitude. Consider an edge $e$ in $\kappa$ with ingoing faces $f_{i,1},\ldots f_{i,n}$, and outgoing faces $f_{o,1},\ldots,f_{o,m}$. Suppose that the associated intertwiner, which is an invariant map
\begin{eqnarray}
\iota_e\;:\;\bigotimes_{I=1}^nV_{j_{i,I}^+,j_{i,I}^-}\;\longrightarrow\;\bigotimes_{J=1}^mV_{j_{o,J}^+,j_{o,J}^-}
\end{eqnarray}

\noindent factorises in the following sense: The incoming and outgoing faces can each be separated into two sets such that there are EPRL intertwiners
\begin{eqnarray*}
\iota_1\;&:&\;\bigotimes_{I=1}^kV_{j_{i,I}^+,j_{i,I}^-}\;\longrightarrow\;\bigotimes_{J=1}^lV_{j_{o,J}^+,j_{o,J}^-}\\[5pt]
\iota_2\;&:&\;\bigotimes_{I=k+1}^nV_{j_{i,I}^+,j_{i,I}^-}\;\longrightarrow\;\bigotimes_{J=l+1}^mV_{j_{o,J}^+,j_{o,J}^-}
\end{eqnarray*}

\noindent such that
\begin{eqnarray}
\iota_e\;=\iota_1\otimes\iota_2.
\end{eqnarray}

\begin{figure}[hbt!]
\begin{center}
    \psfrag{v1}{$v_i$}
    \psfrag{v2}{$v_f$}
    \psfrag{e1}{$\iota_1$}
    \psfrag{e2}{$\iota_2$}
    \psfrag{e}{$\hspace{-25pt}\iota_1\otimes\iota_2$}
    \includegraphics[scale=0.75]{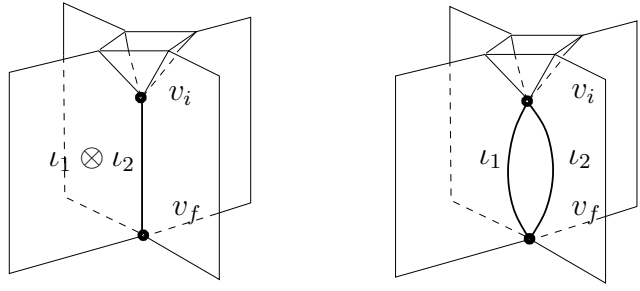}
    \caption{Pulling the edge $e$ with $\iota_e=\iota_1\otimes \iota_2$ apart into two edges $e_1$, $e_2$ with $\iota_1$ on one and $\iota_2$ on the other.}\label{Fig:PullEdge}
\end{center}
\end{figure}

\noindent Then it is possible to deform the two-complex by splitting the edge $e$ into two edges $e_1$ and $e_2$, each of which have the same starting- and ending vertices as $e$ (see figure \ref{Fig:PullEdge}). All the faces $f_{i,I}$, $f_{i,J}$ with $I=1,\ldots, k$, $J=1,\ldots l$ are attached to $e_1$, and those with $I=k+1,\ldots, n$, $J=l+1,\ldots m$ are attached to $e_2$. Note that the resulting spin foam two complex has two edges where the original two-complex had only one, while the number of faces and vertices has not changed. We equip the edge $e_1$ with the intertwiner $\iota_1$, the edge $e_2$ with $\iota_2$, and call the resulting spin foam $\kappa'$.

\noindent Since for two square matrices $A, B$ one has ${\rm tr}(A\otimes B)={\rm tr}(A){\rm tr}(B)$, and furthermore $\iota_e\iota_e^{\dag}=(\iota_1\iota_1^\dag)\otimes(\iota_2\iota_2^\dag)$, we find that
\begin{eqnarray}
\mathcal{A}_{e_1}\mathcal{A}_{e_2}\;&=&\;\frac{1}{{\rm tr}(\iota_1\iota_1^\dag)}\frac{1}{{\rm tr}(\iota_2\iota_2^\dag)}\;=\;\frac{1}{{\rm tr}(\iota_e\iota_e^\dag)}\;=\;\mathcal{A}_e
\end{eqnarray}

\noindent If the beginning- and ending vertex of $e$ are $v_i$ and $v_f$ respectively, then the vertex amplitudes $\mathcal{A}_{v_i}$ and $\mathcal{A}_{v_f}$ are a priori different, because the vertex graphs $\gamma_{v_i}$ and $\gamma_{v_f}$ have changed as well. In both graphs there is a vertex $w_i$ and $w_f$ respectively, which corresponds to the edge $e$. After the "splitting apart", in each vertex graph this vertex $w_{i,f}$ is replaced by two vertices $w_{i,f}^{1}$, $w_{i,f}^{2}$. Hence the "pulling apart" of the edge $e$ in $\kappa$ corresponds to pulling apart the two vertices $w_i$ and $w_f$ in the neighbouring graphs $\gamma_i$ and $\gamma_f$.

\begin{figure}[hbt!]
\begin{center}
    \psfrag{v1}{$w_i$}
    \psfrag{w1}{$w_i^1$}
    \psfrag{w2}{$w_i^2$}
    \psfrag{S}{$S^3$}
    \psfrag{g1}{$\gamma_{v_i}$}
    \includegraphics[scale=0.75]{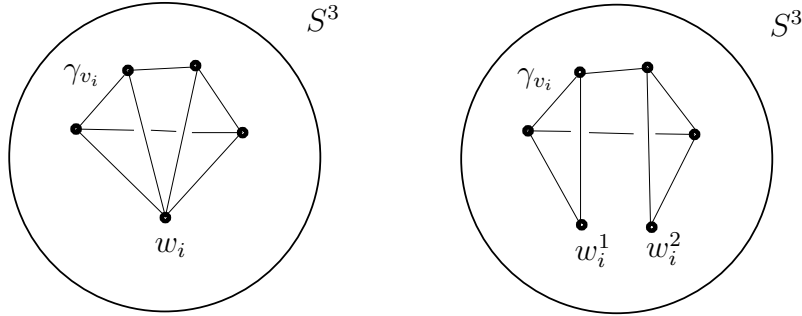}
    \caption{Pulling the edge $e$ apart corresponds to pulling apart the corresponding vertex $w_i$ of the vertex function $\psi_{v_i}$ (and $w_f$ of $\psi_{v_f}$).}\label{Fig:VertexPullEdge}
\end{center}
\end{figure}

\noindent Because of the definition of the vertex amplitude (\ref{Gl:VertexAmplitude}), the vertex amplitudes itself are invariant under this operation, e.g.
\begin{eqnarray}\label{Gl:InvarianceOfVertexAmplitudeUnderPullingEdgesApart}
\mathcal{A}_{v_i}(j_f^\pm,\ldots,\iota_1\otimes\iota_2,\ldots)\;=\;\mathcal{A}_{v_i}(j_f^\pm,\ldots,\iota_1, \iota_2,\ldots)
\end{eqnarray}

\noindent because it is computed simply by contracting the indices of intertwiners on edges meeting in $v_i$. Due to the orientation of the edge $e$, the intertwiner associated to $w_f$ is $\iota_e^\dag=\iota_1^\dag\otimes\iota_2^\dag$, which is also of product form, and hence for $\mathcal{A}_{v_f}$ an equation similar to (\ref{Gl:InvarianceOfVertexAmplitudeUnderPullingEdgesApart}) holds. As a consequence, the whole spin foam amplitude is left invariant under the move described above:
\begin{eqnarray}
Z[\kappa]\;=\;Z[\kappa'].
\end{eqnarray}

\begin{figure}[hbt!]
\begin{center}
    \psfrag{v1}{$w_i$}
    \psfrag{w1}{$w_i^1$}
    \psfrag{w2}{$w_i^2$}
    \psfrag{S}{$S^3$}
    \psfrag{g1}{$\gamma_{v_i}$}
    \includegraphics[scale=0.75]{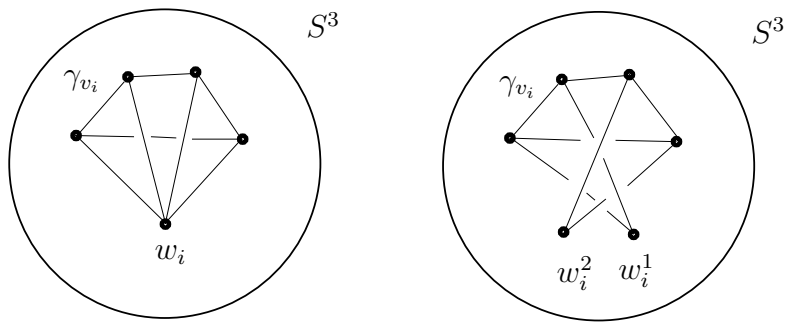}
    \caption{Different way of pulling the vertex $w_i$ apart into two vertices $w_i^1$, $w_i^2$, resulting in a vertex graph with different knotting class in $S^3$.}\label{Fig:VertexPullEdgeAlt}
\end{center}
\end{figure}

\noindent Note that, unlike in three dimensions (as e.g. figure \ref{Fig:PullEdge} suggests), in four dimensions there may be several ways of pulling the edge apart into two edges. This can be seen most easily by realising that there are several ways of e.g. pulling apart the vertex $w_i$ in the vertex graph $\gamma_{v_i}$ into $w_i^1$ and $w_i^2$ in figure \ref{Fig:VertexPullEdge}: Instead of just separating the two vertices to produce a planar $\gamma_{v_i}$, one could have separated $w_i$ into two vertices $w_i^1$ and $w_i^2$, producing a graph that has a nontrivial knotting class (see figure \ref{Fig:VertexPullEdgeAlt}). If one chooses an according way of pulling the vertex $w_f$ apart, this defines a different way of pulling apart the edge $e$. The fact that there are different ways of pulling edges apart differing by knotting classes will play a paramount r\^{o}le later.

\noindent Of course, we allow not only to "pull apart" some of the edges, but also allow for the the inverse move, which corresponds to "merging" two edges $e_1$ and $e_2$ in a spin foam $\kappa$ which have the same beginning- and ending vertices. The resulting edge $e$ then is equipped with the tensor product of the intertwiners $\iota_e:=\iota_1\otimes \iota_2$. The resulting spin foam $\kappa$ then evidently has the same amplitude as $\kappa$, i.e.
\begin{eqnarray*}
Z[\kappa']\;=\;Z[\kappa].
\end{eqnarray*}

\subsection{Consistent deformations as equivalence relation}

The two "moves" (and their inverses) that have been described in the last sections generate, together with trivial subdivisions of faces and edges (and their inverses), an equivalence relation of spin foams $\kappa$. They can be viewed as continuous deformations of the two-complex such that some bits of the two-complex are allowed to be put together or split apart. We will see later how one can use this in order to also allow faces to pass through each other in a controlled way. We term these "consistent deformations". Since the spin foam amplitude $Z[\kappa]$ does not change under a consistent deformation, one can restrict the sum in the physical inner product (\ref{Gl:PhysicalInnerProductWithFoams}) over the set of equivalence classes, where $\kappa\sim\kappa'$ iff $\kappa$ and $\kappa'$ are consistent deformations of each other:
\begin{eqnarray}\nonumber
\langle\psi_f|\psi_i\rangle_{\rm phys}\;&=&\;\sum_{\kappa\,:\,\psi_i\stackrel{\kappa}{\rightarrow}\psi_f}Z[\kappa]\\[5pt]\label{Gl:PhysicalInnerProductRestriced}
&=&\;\sum_{[\kappa]\,:\,\psi_i\stackrel{\kappa}{\rightarrow}\psi_f}Z[\kappa].
\end{eqnarray}

\noindent It should be noted that inequivalent spin foams $\kappa$ might have different "orbit size" under the action of consistent deformations, and one might define the inner product (\ref{Gl:PhysicalInnerProductRestriced}) with an additional factor $F[\kappa]$ depending only on the equivalence class $[\kappa]$, measuring the relative size of these orbits. Different choices for $F$ all result in diffeomorphism-invariant physical inner products, so there is no a priori way of choosing one.\footnote{In defining the spatially-diffeo invariant Hilbert space of canonical LQG, a similar freedom of choosing such a factor exists, see e.g. \cite{DIFF} for a discussion.} Since - even when restricting the sum to equivalence classes under consistent deformations - the sum (\ref{Gl:PhysicalInnerProductRestriced}) diverges anyway, we do not bother too much about the actual choice of $F$, set $F[\kappa]\equiv 1$ from now on, and remark that all the following statements in this article remain valid for "sensible" choices of $F$.

\section{The physical Hilbert space}\label{Sec:PhysHil}

\subsection{The "knotting spin foam"}

In this section we will come to a crucial ingredient for our final result: We will consider two spin network functions $\psi_i$, $\psi_f$ living on graphs $\gamma_i$ and $\gamma_f$ which have the same combinatorics (i.e. adjacency matrix), but different knotting classes. The goal in this section is to compute the spin foam amplitude $Z[\kappa_0]$ for a spin-foam $\kappa_0:\psi_i\to\psi_f$ which mediates between the two, and is essentially "unknotting" $\gamma_i$ into $\gamma_f$. It is not hard to see that all these unknottings can be achieved by a successive application of the following "move":

\begin{figure}[hbt!]
\begin{center}
    \psfrag{A}{\hspace{-10pt}$\longrightarrow$}
    \psfrag{j1}{$k_i$}
    \psfrag{j2}{$k_f$}
    \psfrag{p1}{$\psi_i$}
    \psfrag{p2}{$\psi_f$}
    \includegraphics[scale=0.5]{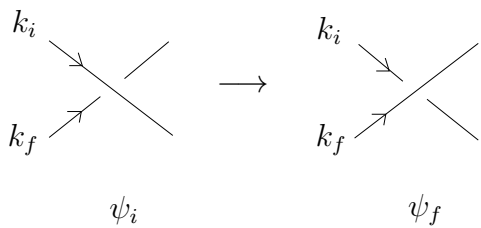}
    \caption{Graphs with different knotting classes can be related to each other by subsequently crossing lines.}\label{Fig:Unknot}
\end{center}
\end{figure}

\begin{figure}[hbt!]
\begin{center}
    \psfrag{k}{$k$}
    \psfrag{l}{$l$}
    \psfrag{v}{$v$}
    \psfrag{i}{$\psi_i$}
    \psfrag{f}{$\psi_f$}
    \includegraphics[scale=0.4]{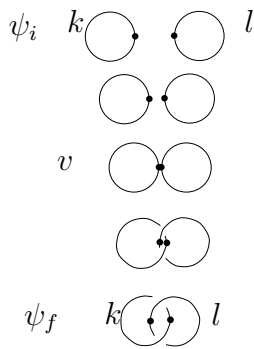}
    \caption{We consider the example of $\psi_i$ consisting of two loops (with spins $k$ and $l$), and $\psi_f$ consisting of the same loops which are linked with each other. }\label{Fig:CD}
\end{center}
\end{figure}

\noindent Without loss of generality we assume that $\psi_i$ and $\psi_f$ are situated on graphs consisting of two loops, where in $\psi_f$ the two loops are linked, while in $\psi_i$ they are not linked. The transition between the two is given by the spin foam $\kappa_0$, which can be described as follows: The two loops start at $\psi_i$ as two linked loops, approach each other, meet at their respective vertex, and pass through each other, eventually ending linked as $\psi_f$ (see figure \ref{Fig:CD}). The resulting spin foam $\kappa_0$ has one internal vertex $v$, which is the point in which the two loops meet.

\noindent We now show that $Z[\kappa_0]=1$, which will be the key point in our analysis. First we note that $\kappa_0$ has two internal faces, four internal edges (each going from one of the four boundary vertices to $v$) and one internal vertex. Using the definition (\ref{Gl:BoundaryAmplitudeSymmetric}), one easily sees that the face amplitudes and the boundary edge amplitudes cancel. Furthermore the boundary vertex amplitudes $\mathcal{B}_v$ cancel two of the four edge amplitudes $\mathcal{A}_e$. If we denote the edges describing the history of the vertices of $\psi_i$ until they meet at $v$ by $e_1$ and $e_2$, we obtain
\begin{eqnarray}\nonumber
Z[\kappa_0]\;&=&\;\mathcal{A}_{e_1}\mathcal{A}_{e_2}\mathcal{A}_v\\[5pt]\label{Gl:KnottingAmplitude1}
&=&\;\frac{1}{(2j_k^++1)(2j_k^-+1)}\frac{1}{(2j_l^++1)(2j_l^-+1)}\mathcal{A}_v
\end{eqnarray}

\noindent where $2j_k^\pm=|1\pm \gamma|k$ and $2j_l^\pm=|1\pm \gamma|l$. Here we have used  (\ref{Gl:EdgeAndVertexIdentity}) and (\ref{Gl:BoundaryAmplitudeSymmetric}), remembering that the edges $e_1,\,e_2$ carry the identity intertwiners.\footnote{To be precise, since the intertwiners on the boundary vertices carry the identity intertwiners, e.g. $\hat\iota={\rm id}_{V_k}$, the edges of $\kappa$ carry $\iota=\phi(\hat\iota)=\frac{2k+1}{(2j_k^++1)(2j_k^-+1)}{\rm id}_{V_{j_k^+,j_k^-}}$, which is a nontrivial multiple of the identity intertwiner. However, due to the definition (\ref{Gl:EdgeAndVertexIdentity}) of the edge amplitude, the spin foam is not sensitive to this factor.}

\begin{figure}[hbt!]
\begin{center}
    \psfrag{k1}{$\!\!j_k^\pm$}
    \psfrag{k2}{$\!\!j_l^\pm$}
    \psfrag{S}{$S^3$}
    \psfrag{g}{$\gamma_v$}
    \includegraphics[scale=0.5]{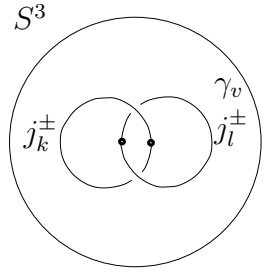}
    \caption{The neighbouring spin network, situated on the graph $\gamma_v=S^3\cap \kappa_0$. }\label{Fig:Knot}
\end{center}
\end{figure}

All that remains to be done now is to compute the vertex amplitude for the one internal vertex $v$ in $\kappa_0$, which is the point where the two edges are crossing each other in order to "entangle the knot". In order to do so, one needs to intersect the spin foam with a small three-sphere around the vertex $v$, to obtain a graph $\gamma_v$ embedded in the three-sphere. It is not hard to see that $\gamma_v$ consists of two links, i.e. two loops, each with two vertices and two edges (see figure \ref{Fig:Knot})

\noindent The evaluation of this spin network function\footnote{Which does not notice the fact that the two loops form a nontrivial knot - which is the central reason for the results of this article.} can then be done, and one obtains
\begin{eqnarray}
\mathcal{A}_v\;=\;(2j_k^++1)(2j_k^-+1)(2j_l^++1)(2j_l^-+1)
\end{eqnarray}

\noindent which, with (\ref{Gl:KnottingAmplitude1}), results in
\begin{eqnarray}\label{Gl:KnottingAmplitude2}
Z[\kappa_0]\;=\;1
\end{eqnarray}

\subsection{The spin foam sum}

We have shown that the amplitude for the "unknotting" spin foam $\kappa_0:\psi_i\to\psi_f$ is $Z[\kappa_0]=1$, which is equal to the trivial amplitude. We now show that the spin foam $\kappa_0\kappa_0^{-1}:\psi_i\to\psi_i$ is a consistent deformation of the identity foam ${\rm id}:\psi_i\to\psi_i$. This can be seen as follows:

\begin{figure}[hbt!]
\begin{center}
    \psfrag{R}{$\!\!\Leftrightarrow$}
    \psfrag{k}{$\kappa_0\kappa_0^{-1}$}
    \psfrag{i}{${\rm id}$}
    \psfrag{u}{$\Leftrightarrow$}
    \psfrag{d}{$\Leftrightarrow$}
    \includegraphics[scale=0.5]{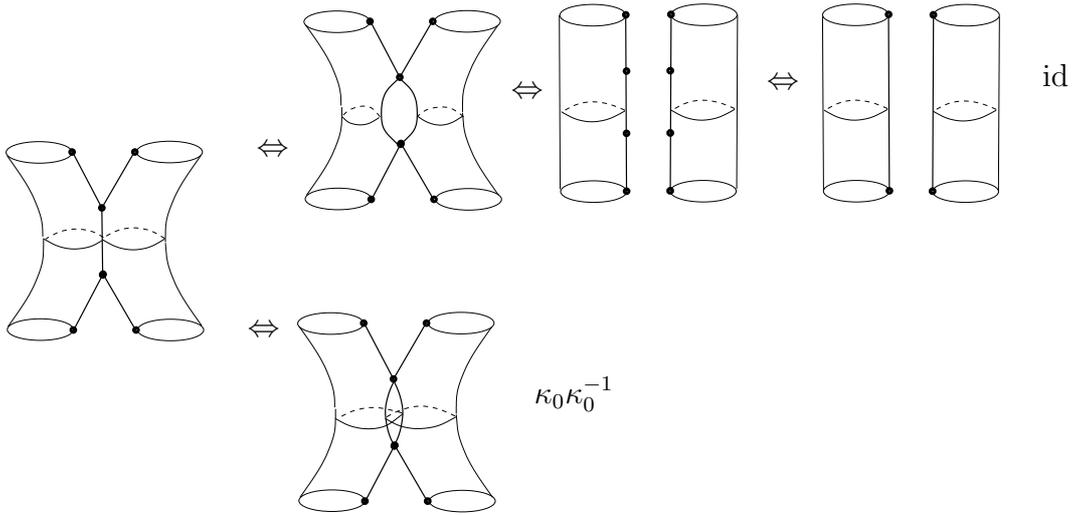}
    \caption{The spin foams ${\rm id}$ and $\kappa_0\kappa_0^{-1}$ can be consistently deformed into each other by application of the pulling moves described above (and their inverses), and removind/adding trivial vertices.}\label{Fig:CD2}
\end{center}
\end{figure}

Starting from the spin foam furthest to the left in figure \ref{Fig:CD2}, let the associated intertwiner to the edge between the two internal vertices be $\iota={\iota}_{1}\otimes {\iota}_{2}$, where $\iota_1$ is the identity map on $V_{j_k^\pm}$, and $\iota_2$ the identity map on $V_{j_l^\pm}$. As discussed in the previous section, there are two ways of pulling the internal edge connecting the two internal vertices apart, differing by a knotting class of the resulting vertex amplitudes. One results immediately in $\kappa_0\kappa_0^{-1}:\psi_i\to\psi_i$, as one can readily see. The other one results in two "tubes" (the histories of the two loops) which are connected at the two internal vertices. Pulling the foam apart at these two vertices and then removing the two vertices (which have become trivial), one arrives at the identity foam ${\rm id}\,:\psi_i\to\psi_i$

As a result, for any kinematical state $\varphi$ we can write down a bijection $\Phi$ between the set of spin foams $\kappa:\varphi\to\psi_i$  modulo consistent deformations and spin foams $\kappa':\varphi\to\psi_f$ modulo consistent deformations by
\begin{eqnarray}\nonumber
\Phi\;:\;[\kappa]\;&\longmapsto&\;[\kappa\kappa_0]\\[5pt]
\Phi^{-1}\;:\;[\kappa']\;&\longmapsto&\;[\kappa'\kappa_0^{-1}]
\end{eqnarray}

\noindent That $\Phi$ is not only well-defined but also a bijection follows from $\kappa_0\kappa_0^{-1}\sim{\rm id}$, i.e. the two spin foams are consistent deformations of each other. With the functorial property (\ref{Gl:MultiplicativeAmplitude}) and (\ref{Gl:KnottingAmplitude2}) one finds that
\begin{eqnarray}
Z[\kappa]\;=\;Z[\Phi\kappa]
\end{eqnarray}

\noindent Therefore, for each (equivalence class of a) spin foam $\kappa:\varphi\to\psi_f$ there is exactly one (equivalence class of a) spin foam $\kappa':\varphi\to\psi_i$, and the spin foam amplitude is the same for each of the two. It follows that

\begin{eqnarray}\nonumber
\langle\psi_i|\varphi\rangle_{\rm phys}\;&=&\;\sum_{[\kappa]:\varphi\to\psi_i}Z[\kappa]\;=\;\sum_{\Phi[\kappa]:\varphi\to\psi_f}Z[\kappa]\\[5pt]\label{Gl:CoreCalculation}
&=&\;\sum_{\Phi[\kappa]:\varphi\to\psi_f}Z[\Phi\kappa]\;=\;\sum_{\kappa':\varphi\to\psi_f}Z[\kappa']\\[5pt]\nonumber
&=&\;\langle\psi_f|\varphi\rangle_{\rm phys}
\end{eqnarray}

\noindent It should be noted that this property follows entirely from the symmetries of the amplitude $Z[\kappa]$, and is independent of the actual way in which the vastly sum $\sum_\kappa$ over spin foams is realised, as long as it is realised such that the symmetries of $Z[\kappa]$ are respected.

Since $\varphi$ was arbitrary, from (\ref{Gl:CoreCalculation}) it follows immediately that the difference $\psi_i-\psi_f$ has zero physical norm, i.e. that $\psi_i$ and $\psi_f$ are mapped to the same physical state in the sense of the rigging map procedure (\ref{Gl:RiggingMap}).

The calculations presented here were for the case of two loops interlinking each other. It should, however, have become clear that the calculations for a generic case of two graphs $\gamma_i$, $\gamma_f$ with the same combinatorics but different knotting classes is performed exactly the same way. All one needs to do is trivially subdividing each pair of edges that shall "pass through" each other during the evolution described by the spin foam with a trivial vertex each. As a result (\ref{Gl:CoreCalculation}) also holds for states $\psi_i$, $\psi_f$ situated on $\gamma_i$, $\gamma_f$. Hence, the states in the physical Hilbert space $\mathcal{H}_{\rm phys}$ do not contain any knotting classes of graphs.

\section{Summary and Discussion}

In this article we have investigated properties of the physical Hilbert space $\mathcal{H}_{\rm phys}$ as defined by the sum-over-spin foams procedure using the Euclidean EPRL amplitude. The result was that $\mathcal{H}_{\rm phys}$ does not contain any knotting information of the graphs in the following sense: The bona fide projector $\eta:\mathcal{H}_{\rm kin}\to\mathcal{H}_{\rm phys}$ maps states that have the same combinatorics but different knotting classes to the same physical state.

This result depends on several assumptions. The most visible assumption is that of the face- and edge amplitudes (\ref{Gl:EdgeAndVertexIdentity}) and (\ref{Gl:FaceAndEdgeIdentity}), as well as the boundary amplitudes (\ref{Gl:BoundaryAmplitudeSymmetric}). This choice was a natural consequence of demanding the invariance of the spin foam amplitude $Z[\kappa]$ under trivial subdivisions (figures \ref{Fig:SubdivisionEdge} and \ref{Fig:SubdivisionFace}), and the product property (\ref{Gl:MultiplicativeAmplitude}). The invariance of $Z[\kappa]$ under subdividing an edge with a vertex (without conditions on the vertex, since the vertices do not carry any data) does, strictly speaking, not follow from the canonical framework. However, it appears to be natural in the light of interpreting a spin foam as a history of a spin network. If one believes that the whole dynamical information is captured by the evolution of the graph and the representation data distributed on it, then that dynamical information is not changed under subdividing an edge with a vertex. Hence the amplitude $Z[\kappa]$ should not be sensitive to this change.

The result we have obtained rested on a special property of the EPRL amplitude, which leads to an invariance of the spin foam amplitude $Z[\kappa]$ under a "consistent deformation" of $\kappa$. It should be noted that we do not propose such a consistent deformation to correspond to a physical symmetry, or be related to diffeomorphisms (other than diffeomorphisms being a subset of the consistent deformations). In particular, the consistent deformations do not form a group in a straightforward way. Nevertheless, the invariance is a fact, and it allows to severely reduce the summation over all spin foams to all spin foams modulo consistent deformations. Hence, as a convenient tool it allows to prove statements about the physical Hilbert space.

The central reason for the independence of the physical Hilbert space of knottings is that the "unknotting" spin foam $\kappa_0$ has unit amplitude, i.e. $Z[\kappa_0]=1$, and the ultimate reason for this is the definition of the vertex amplitude itself. The value of $\mathcal{A}_v$ depends only on the evaluation of a spin network embedded in $S^3$, and this evaluation is independent of the knotting of the embedded graph. It is just sensitive to its combinatorics. It should be noted that this depends of the generalisation to the EPRL amplitude given in \cite{KKL}, which was until then only defined for the case in which the corresponding graph in $S^3$ had a trivial knotting anyway. It appears to be possible to change the generalisation of the amplitude from \cite{EPRL} to general spin foams by changing its sensitivity to the knotting of the vertex graph, e.g. by multiplying the  vertex amplitude by a nontrivial function of knotting invariants, or by equipping the vertices of the spin foam itself with "knotting charges", which, for the sake of consistency, should be trivial in the case of the four-simplex amplitude.

Although the physical Hilbert space does not contain any knotting information of the graphs, it should be emphasized that this does \emph{not} mean that the theory is insensitive to knotting within the space-time four-manifold $\mathcal{M}=\Sigma\times[0,1]$! The latter is a question concerning the expectation values of some (four-dimensional) observables which are computed in the theory. The shape of the boundary Hilbert space is not necessarily making any statements of what possible expectation values arise in the theory, and whether they differ for different observables (e.g. corresponding to differently knotted surfaces in $\mathcal{M}$) or not. A prominent example which illustrates this is three-dimensional gravity as described by the Ponzano Regge model. The boundary Hilbert space, in the Loop formulation, is given by planar spin networks embedded in the two-dimensional boundary manifold \cite{2DLQG}, and therefore contains no knotting information, not even on the kinematical level. The expectation value of observables corresponding to curves embedded in the three manifold (e.g. Wilson lines) however \emph{do} depend on the knotting of the embedding \cite{PR}, being related to well-known knot invariants. It is very well possible that this also happens in the four-dimensional theory. Before one has made a more precise sense of the sum-over-spin foams and has a good idea of what observables in the theory should be, it will nevertheless be hard to make any certain statements about this issue.

\section*{Acknowledgements}

It is a pleasure to thank Jurek Lewandowski for invitations to Warsaw, and him, Wojtiech Kaminski and Frank Hellmann for enlightening discussions. The author furthermore would like to thank Bianca Dittrich for urging him to write this article.

\end{document}